\documentclass[11pt]{article}
\usepackage{latexsym,amssymb}
\usepackage{hyperref}
\usepackage{graphicx}
\begin{document}

\title{
    \bfseries \Large
    Detection of Weak~Gravitational~Waves\\by Interferometric
    Methods\\and Problem of Invertible Calculations}

\author{
    {\scshape \normalsize
    A.\,I.~Golovashkin, G.\,V.~Kuleshova, A.\,M.~Tzhovrebov}\\
    {\itshape \normalsize
    P.\,N.~Lebedev Physics Institute, Moscow} \vspace{6pt}\\
    {\scshape \normalsize
    G.\,N.~Izmailov, Tran~Quoc~Khanh}\\
    {\itshape \normalsize
    Moscow Aviation Institute (a State Technical University)}}

\date{}
\maketitle

\begin{abstract}
    \noindent The fundamental features of the detection of
    non-stationary undulatory perturbations of metrics based on the
    interference effects are considered. The advantage of the
    Aharonov-Bohm effect in superconductors for these purposes in
    comparison with the ordinary optical interference is demonstrated.
    Some circuitries of the interferometric detectors in order to be
    used with SQUID are suggested. The possibilities of lowering the
    noise temperature of the ultraweak signals detectors based on the
    analogy between the processes of high-sensitive measurements and
    the reversible calculations are discussed.
\end{abstract}

\parskip=4pt

\noindent The question about the detectability and the possibility
of principle to carry out in practice the recording of
gravitational waves (GW) had been brought up for the first time in
the works of Bondi~\cite{Bon} and Weber~\cite{Web1}. The Bondi's
mental experiment (friction of the beads that are ``pushed away by
the perturbation of metric'' under the action of GW, a prototype
of the modern laser-interferometric detectors~\cite{Bag,Col}), as
well as the Weber's real experiments~\cite{Sin} with the massive
aluminum antenna, equipped with piezo-sensors, have implied an
energy transfer of GW to the mechanical system. Before these
authors represented their own statements, it has been proposed a
recording method of GW~\cite{JETP}, assuming the conversion of the
wave energy into the elastic-strain energy of a body with
magnetostrictive properties (fig.~1). The possibility of recording
a magnetic response, caused by the deformation of a
magnetostrictive sample, of the superconductive quantum
interferometer (SQUID) has brought the high efficiency to the
method. On the present-day sensitivity the SQUID's are capable to
register \ $10^{-7}\Phi_0\,Hz^{-1/2}$, \ where \ $\Phi_0 \simeq
2.07\!\times\!10^{-15}\,Wb$ \ is the flux quantum. The basic
estimations on the basis of the actual parameters of
magnetostrictive materials have allowed us to think of the
possibility of enhancing the sensitivity of the proposed method in
terms of GW metric tensor oscillation amplitude to the level of \
$|\,\delta g_{ij}\,| \simeq 2.5\!\times\!10^{-23}\,Hz^{-1/2}$.

\vspace{36pt}
\begin{center}
  \includegraphics[width=129mm,keepaspectratio]{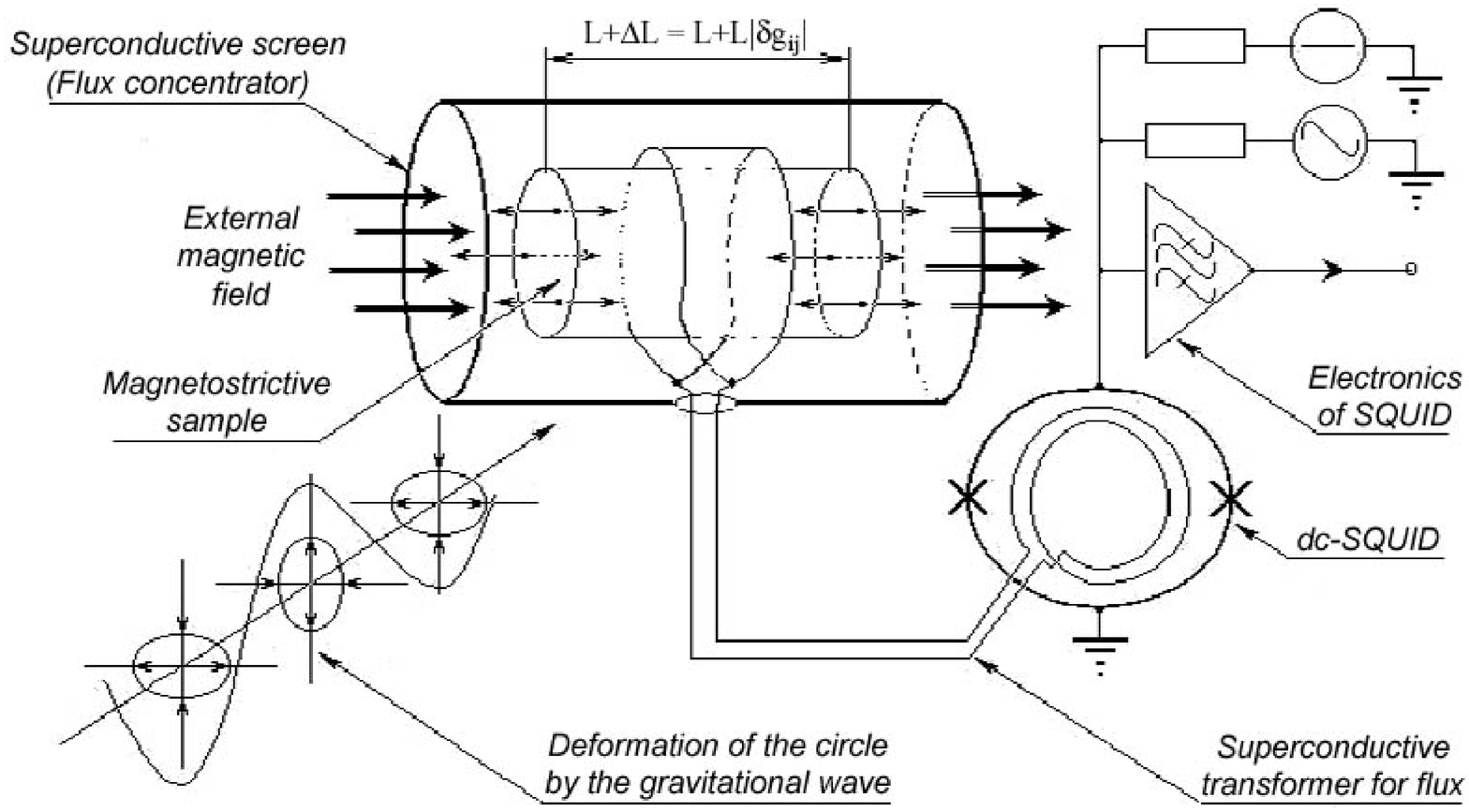}\par
  \vspace{24pt}
  \parbox{117mm}{\small {\itshape Figure 1.} A method for the detection of GW that
    transforms the energy of GW into the energy of elastic deformation
    of a magnetostrictive sample.\par
    \vspace{4pt}
    \parindent=5mm The response from the magnetostrictor to the flux: $\Delta\Phi =
    S\Lambda E|\Delta L/L|$, where $S = 200\,cm^2 = 2\times10^{-2}\,m^2$ is the area
    of the base, $\Lambda\gtrsim2\times10^{-9}\,T/Pa$ is the sensitivity,
    and $E = 200\,GPa$ is the Young modulus of the magnetostrictor.\par
    \vspace{4pt}
    \parindent=5mm The resolution capacity of SQUID: $\delta\Phi = 10^{-7}\Phi_0\,(1/\sqrt{Hz}) =
    2.07\times10^{-22}\,Wb/\sqrt{Hz}$. As $\Delta\Phi$ is estimated by $\delta\Phi$ $\Rightarrow$
    the sensitivity of the magnetostrictor in terms of the metric tensor variational amplitude:
    $|\delta g_{ij}| = \Delta L/L = 2.5\times10^{-23}\,(1/\sqrt{Hz})$.}
\end{center}
\vspace{36pt}

However, when considering the recording processes based on the
energy conversion of a GW field into the measurable signals, it
must be taken into account that one out of the profound problems
of the Einstein general relativity theory is the problem of
determining the energy of the gravitational field
itself~\cite{Pau,Web2}. In particular, this problem comes up in
the calculation of total energy fluxes transported by GW from a
source. The divergences, arising at the same time about the
simplest symmetries of the problem, have forced some
authors~\cite{Mul} to deduce that GW, being purely ``geometrical
objects'' themselves, in general do not transfer the energy. It is
possible that the problems with the energy in physics of the
nonstationary gravitation themselves are essentially original
failures of the experimental detection of the gravitational
radiation by the traditional methods. If this is really the case,
then GW should be searched as nonstationary variations of metric
by the direct approaches without transforming them into the
vibrational energy of probe elastic bodies~\cite{Sin,WWW} or the
oscillations of the optical interferometer's
mirror~\cite{Bag,Col}. The direct measurement of the metric's
variations in interferometric experiments is possible due to
measuring changes of the optical path difference, caused by the
space-time curvature under the action of undulatory gravitational
perturbations, in the interferometer's arms. The arising phase
difference periodically shifts the interference pattern, that
leads to the change of a light intensity at the recording
photomultiplier's input, and can be estimated in the end from the
formula \ $\delta \varphi = {2\pi L|\,\delta
g_{ij}\,|}/{\lambda}$, \ where $L$ is the interferometer's base
length, \ $\delta g_{ij}$ \ is the metric tensor's variation,
$\lambda$ is the operating wavelength of the interferometer. It is
clear that in such experiments one should use in the
interferometer the highly monochromatic light, because the
monochromaticity holds down the error of the phase measurement.
The radiation of ultrastable lasers meets this requirement.
Because so far the X-ray lasers are not  yet created and
ultraviolet ones are not yet ultrastable, then the operating
wavelength is still restricted below in the visible and near IR
ranges (e.\,g., the high-stable infrared line of He-Ne
laser~\cite{Bag}).

Though it is possible to reduce the operating wavelength (in order
to increase the phase response capacity of an interferometric
experimental system) by taking advantage of the quantum
interference effect in weak coupling superconductors. The
effective wavelength of the Cooper pairs' condensate,
corresponding to the quantum interference in the geometry of the
Aharonov-Bohm effect, is represented by formula $\lambda_C = \pi
\hslash /(eA)$. For comparison with optics it is possible to write
down the numerical value of the coefficient that relates
$\lambda_C$, expressed in {\it angstrom} ($\mathring{A}$), to the
modulus of vector potential $\mathbf{A}$, expressed in {\it
tesla$\times$metre} ($T\,m$): \ $\lambda_C\,[\mathring{A}] \approx
10^{-7}A^{-1}\,[T\,m]$. \ Consequently, even the technically
attainable weak fields (with the orders of \ $A \simeq
10^{-6}\,T\,m = 1\,\textit{\OE}\,cm$, \ $\lambda_C \simeq
0.1\,\mathring{A}\,$) \ make the quantum interference of
superconductive condensate (i.\,e. the Josephson effect and the
Aharonov-Bohm effect) more preferred than optical interference on
the terms of operating wavelength in experiments to detect the
gravitational perturbations (when the base $L$ of the
interferometer is considerable). In order to unbind the
interference systems from the restriction on their sensitivity, it
is necessary to use the light that has not just the high
monochromaticity and stability, but also the considerable power
(about hundreds of {\it watts}) at the way in of the
interferometers~\cite{Col}. The last condition is imposed on the
photoelectric multiplier, which elsewise will be unable accumulate
any signal with the sufficient value of  the signal/noise ratio
even in the single-quantum counting regime. Under the very
conditions of superconductivity the stability of parameters is
ensured comparatively simply by freezing the magnetic flux, and in
this case a critical current of high intensity is provided instead
of an optical radiation of the required great power.

The principle of GW detection without converting the energy of
waves into the oscillation energy of elastic samples in a certain
sense is related to the issue on the possibility of information
transfer without transferring energy. In our case, on the one
hand, the problem to so general extent is not the issue, because
the interferometric detection methods allow the indirect energy
exchange. In interferometric detectors the GW ``control'' the
phase. Being by itself nonlinear, such an effect permits the
response energy to exceed the energy of the initiating
perturbation. However, on the other hand, and in the most general
formulation, the problem of transferring information without
transferring energy obviously permits an affirmative answer that
involves directly the interferometric detection. This answer
should be searched in the theory of reversible calculations on a
quantum computer. As it is known, the reversibility on a quantum
level~\cite{Fey} is made available by the fact that in the course
of calculation the states are always transformed while they are
still the proper states regarding the initial Hamiltonian of the
problem. In this case the states may be degenerated according to
energy.

In that way, the interference methods of detecting nonstationary
variations of the metric appear, by implication, to be concerned
with the theory of quantum computing. However, in this case it
rather takes place the closer relation with namely reversible
computations. In fact, the customary proposition, claiming that in
order to process one bit of information it is needed to dissipate
no less than $kT$ of energy, refers to only the irreversible
calculations. In reversible computations the ``redundant'' entropy
is not produced.  In this case, even in irreversible computations
during the transformations the energy of the initial and final
states of the system usually is still the same (two equal
energetic minimum, separated by a potential barrier), but the
dissipation of energy and/or the entropy growth is caused by the
irreversibility of any transition from some initial state to a
specified final state. It is clear that regarding such a
transition a one-valued description of the inverse process is
impossible indeed.

In essence, the simplest circuit~\cite{Ben} of reversible
computations implies simply the preservation, during the
computation, of all source data. It enables the computation to
transform these data at any step, but the reversibility of
calculation allows to exclude the dissipation. If to draw an
analogy of the computations with measurements, then the main
source of entropy, which is the necessities of prearranging at
first a computer in the specified initial state, corresponds to an
absorption of an idle frequency in super low-noise parametric
amplifiers~\cite{Rob}. It is the inconvertibility of absorptive
process of idle frequency results in the dissipation in parametric
systems and does not allow to obtain here the ``absolute zero''
noise temperature. The generalization of the concept of
transformation reversibility of system states simulating the
computational process to a quantum level by the strategy
construction of calculation is given, excluding the reduction of a
wave function . Such is possible under transformations brought to
operator activity on its eigenstates. In this case the uncertainty
of eigenvalues becomes zero. Latter allows to draw an analogy of
calculations on a quantum computer with measurements in parametric
systems with a quantum squeezing.

More generally, the computation reversibility is achieved by the
construction of such algorithm, when the column vector readings of
computation result \ $\vec{r}$ \ is received from the vector of
input data \ $\vec{e}$ \ via the transforming matrix \ $\hat{M}$,
\ tolerating the construction of the inverse matrix \
$\hat{M}^{-1}$, \ so \ $\vec{r} = \hat{M}\,\vec{e}$, \ and/or \
$\vec{e} = \hat{M}^{-1}\,\vec{r}$. \ The reversible algorithms
allow to produce the information processing without the increase
of an entropy as of information, as thermodynamic. This means that
on processing of one information bit is not required to diffuse
out the $kT$ amount of energy. In this case it is possible to
believe that the effective temperature of a computer is equal to
zero.

\vspace{18pt}
\begin{center}
  \includegraphics[width=117mm,keepaspectratio]{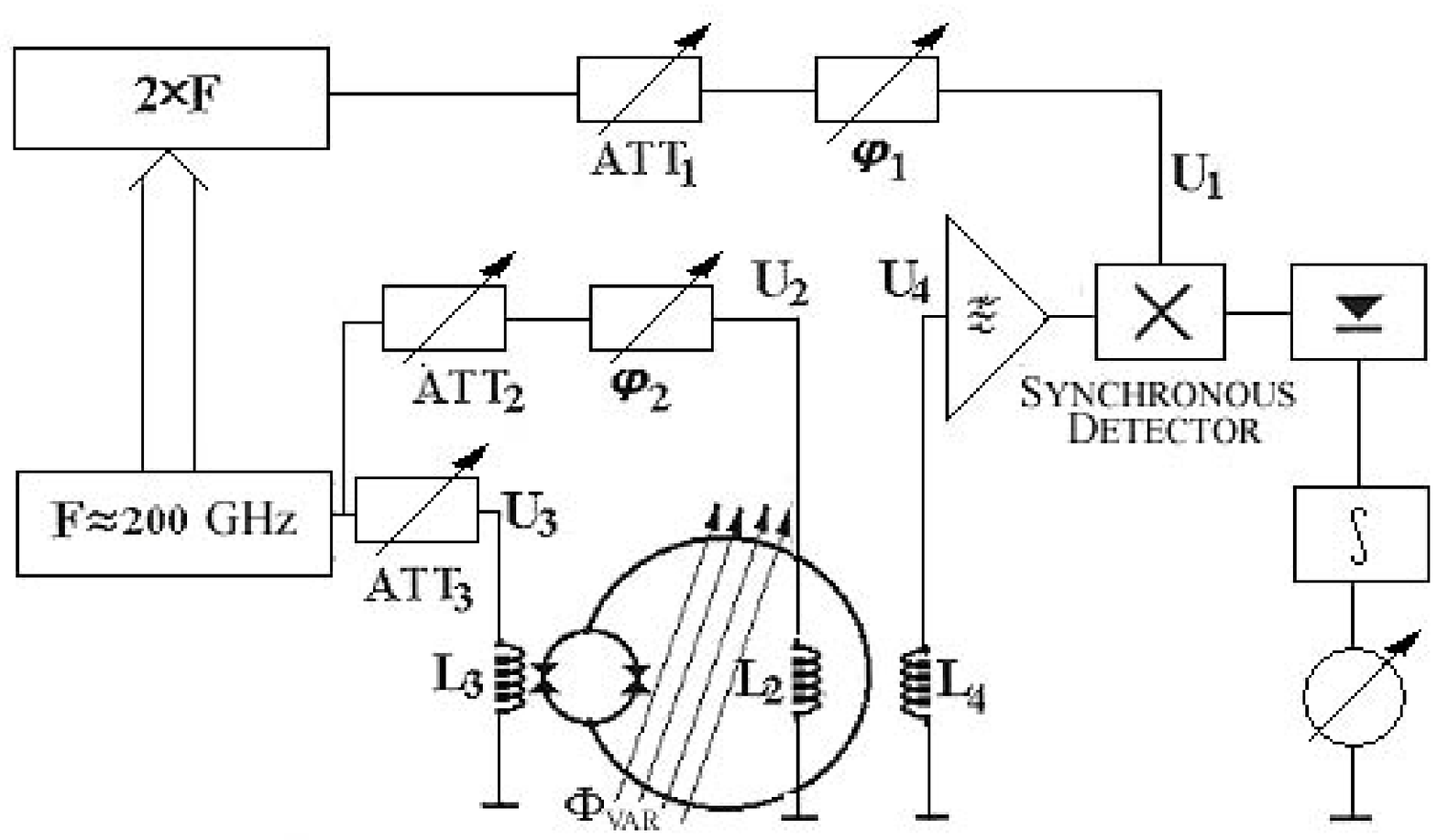}\par
  \vspace{6pt}
  \parbox{117mm}{\small {\itshape Figure 2.} The circuit diagram of a dynamic squeezing anhysteresis
  UHF-SQUID.}
\end{center}
\vspace{36pt}

It is apparent from the fluctuation-dissipative theorem, in the
measurement process the final noise temperature reveals because of
the nonzero imaginary part of ageneralized susceptibility of a
system, that is due to the finite input resistance of a receiver.
Namely the active resistance is responsible for the dissipation of
the energy in measurements. This value in the full sense
characterizes the irreversibility of the measurement process. The
parallel between the computations and the measurements points of
view points here on an advance possibility of the zero noise
temperature, in case of possiblities to construct the algorithm
reversible measurements. Clearly, such measuring receiver must be
designed from pure-jet nonlinear elements. The parametric
amplifier or the anhysteretic HF-SQUID's are the best on this
role. In fig.~2 the flow-chart of the anhysteretic HF-SQUID with a
quantum squeezing is given.  The effect of which has been
previously considered by authors of this message in the
work~\cite{Gol}. In anhysteretic SQUID the role of nonlinear
reactivity is played by the kinematic inductance of Josephson
tunnel junction, under control of the outer magnetic flux,
introduced in superconductive ring of SQUID. As an illustration of
possible approaches on the creating of the technique reversible
measurements let us consider an activity of the simplest
three-frequency nondegenerate paramplifier with a nonlinear
capacitance. The signal amplification here is achieved due to an
insertion of a negative active component of an impedance on an
input frequency. In this case the negative active impedance arises
solely due to mixing of three frequencies (the input, the idle and
the pumping) on a nonlinear capacitance (fig.~3). Therefore,
negative active component is obtained as a result of the work of
purely reactive elements. At first glance, this system is not to
comprise active resistances and according to the
fluctuation-dissipative theorem may be characterized by the zero
noise temperature. However, the parametric amplifier is not yet
intended as the absolutely reversible measuring device. For its
operation the absorbing load of the idler frequency is necessary.
This resistive load makes different the noise temperature from
zero. Consequently, the designing problem of parametric amplifier
with the zero noise temperature or close to one is brought to the
problem of development of the reversible dissipationless load.

\vspace{36pt}
\begin{center}
  \includegraphics[width=129mm,keepaspectratio]{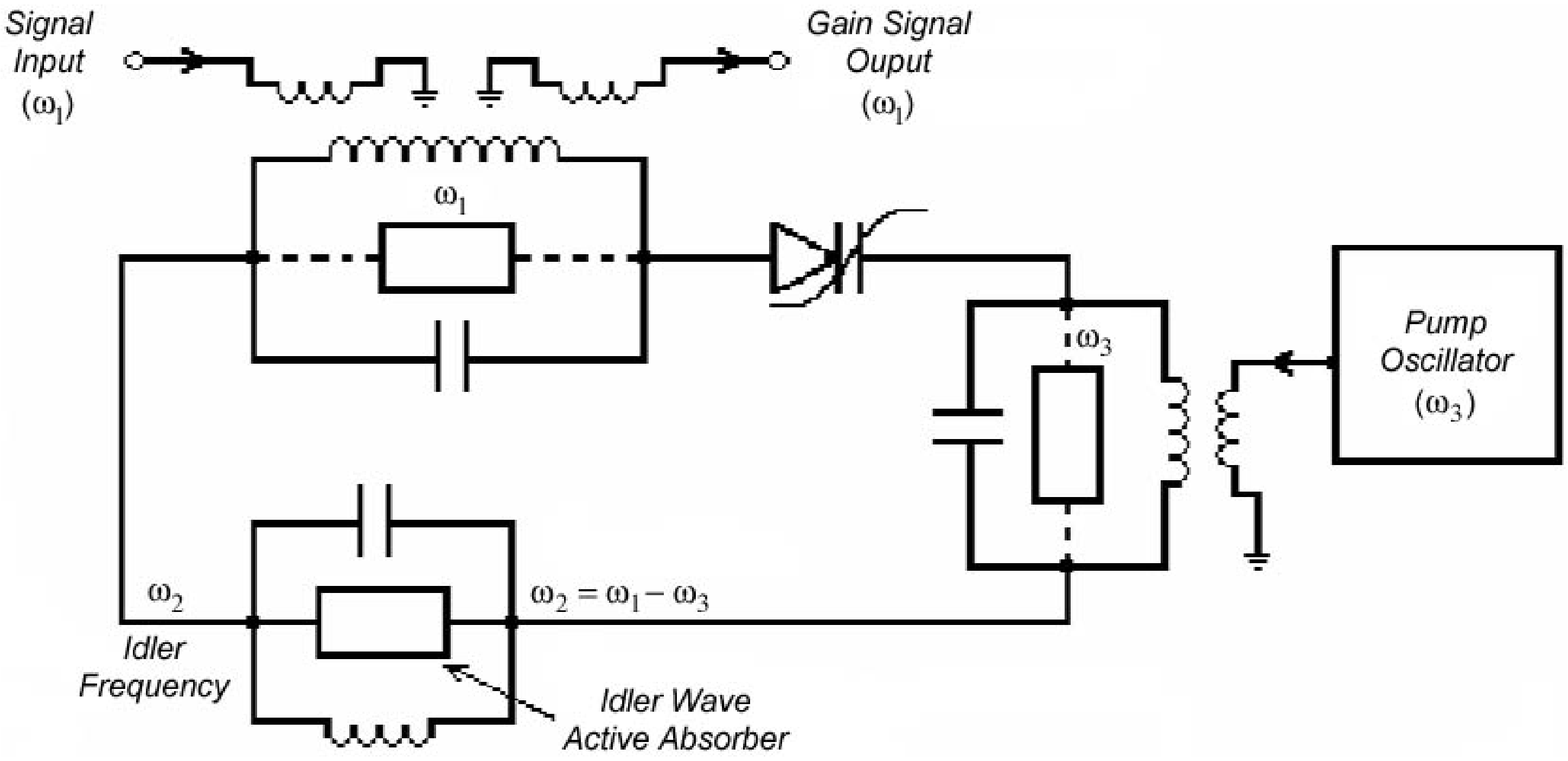}\par
  \vspace{24pt}
  \parbox{117mm}{\small {\itshape Figure 3.} The schematic diagram
  of a simplest three-frequency non-degenerate parametric amplifier
  with a non-linear capacitance.}
\end{center}
\vspace{36pt}

Here are three possible routes to solve the problem. The first one
is analogous to the approach to reversible computations. In order
to turn the process of information processing it is necessary to
all the time to retain results of intermediate calculations. The
latest asks the sufficient volume of the main memory. In this case
one can produce only until for the storage of intermediate data
memory there is an open place. Analogously, during the limited
time, while in high-quality resonator does not end the transition
processes, the initial processing can be regarded as
dissipationless (or, more precisely, low dissipative) loading,
while considering the Universe as an infinite-mode resonator
without any time restriction needed for establishing the balance.
On the other hand, it is possible simply to believe that the
Universe has the quite low noise temperature as well as the load
has. Indeed, until the forcing equilibrium has not been made, the
energy absorbed by the resonator is mainly putting in the growth
of an oscillation amplitude, and it is not just dissipates in the
form of the loss compensation for an oscillation period. Rather
than to absorb the energy of an idler frequency in a resonator in
the nonstationary regime, it could be radiate in outer space also,
considering the Universe as the resonator with neither than
unlimited time of equilibrium setting. On the other hand it is
possible to simply believe that the Universe is a load with the
rather low noise temperature. The second path of the
dissipationless absorption is possible due to the reversibility
property of a parametric amplifier itself. According to Manly-Row
formula, it is easy to income in the condition of attenuation of
input signal instead of amplification by means of the selection in
a proper way of the relation between the working, the idle and
pump frequencies. In such way, the parametric amplifier will be
converted into the  parametric load. As a result the circuit of
the idler frequency of the fundamental amplifier must be linked
with the input circuit of a parametric load. Such load is not
intended as an absolutely dissipationless, because there must be
the absorber ``of its'' idle frequency in the reversed amplifier.
However, the noise temperature of a parametric load can be done
below the physical temperature of the absorber. For a rough
estimation of possiblities to lower one it is necessary to take
the ratio of an input impedance of the parametric load and the
impedance of the proper absorber of the idle frequency load. In
this case the input impedance as a result of a corresponding
circuit adjustment may be done however large. Really, one  is a
relation of an input voltage variation, corresponding to the
increment of an input current, to the value of latter.  The
variation of an input voltage as the response of a nonlinear
system on the current influence may be ``made'' the many times
stronger than in linear case. On the role of the ``third route''
pretends the possibility of an interferometric suppression of a
signal of the idler frequency (here is possible the analogy as
with principle of coated optics, and with quantum computations
too). For implementing of interferometric supressing the input
signal needs to give on two equal parametric amplifiers with the
combined pumping generator and inductively coupled loopes of the
idler frequency, which are geometrically located so that the
magnetic fluxes in them became mutually antiphase.

It is obviously the use of principles of state squeezing in the
quantum interferometer, recording the nonstationary variations of
metrics, would allow us to increase considerably the receiver
sensitivity of a GW signal. However this can be realized only
after following upgrade of a measurement sensitivity on the basic
principle of designing of the strategy of classical reversible
measurements with zero (or vanishing) noise temperature. If to
base on the finiteness of the energy flow rate across unit area by
the transported GW and to use for its estimation the conventional
formula of general relativity
$$S_G = c\,W_G = {{c\,^3\,\omega\,^2}\over16\,\pi\gamma}\cdot(h_{yy}^2\,+\,h_{yz}^2),$$
when it is clear the process of the energy transfer between a wave
and the probe body of a classical receiver becomes
``superunreversible''. Indeed, it is easy to compare the energy
density in gravitational waves,
$$W_G = {{c\,^2\,\omega\,^2}\over16\,\pi\gamma}\cdot(h_{yy}^2\,+\,h_{yz}^2),$$
and in acoustical ones, $$W\!_{Aq} = E\,\varepsilon\,^2,$$ when
the variation amplitude of the metric tensor \ $h_{ij}$ \ is equal
to the amplitude of the specific elongation (strain tensor \
$e_{ij}$, \ Young modulus $E \approx 5\,GPa$) of the elastic
medium. At the frequency $1\,kHz$ the ratio of energy density for
these waves ($W_G/W_{Aq}$) appears to be enormous, approximately
$10^{34}$. It turns out that the empty Euclidean space possesses
the huge elasticity; the effective Young modulus by the order of
34 is higher than at a ``usual'' matter.

\vspace{18pt}
\begin{center}
  \includegraphics[width=117mm,keepaspectratio]{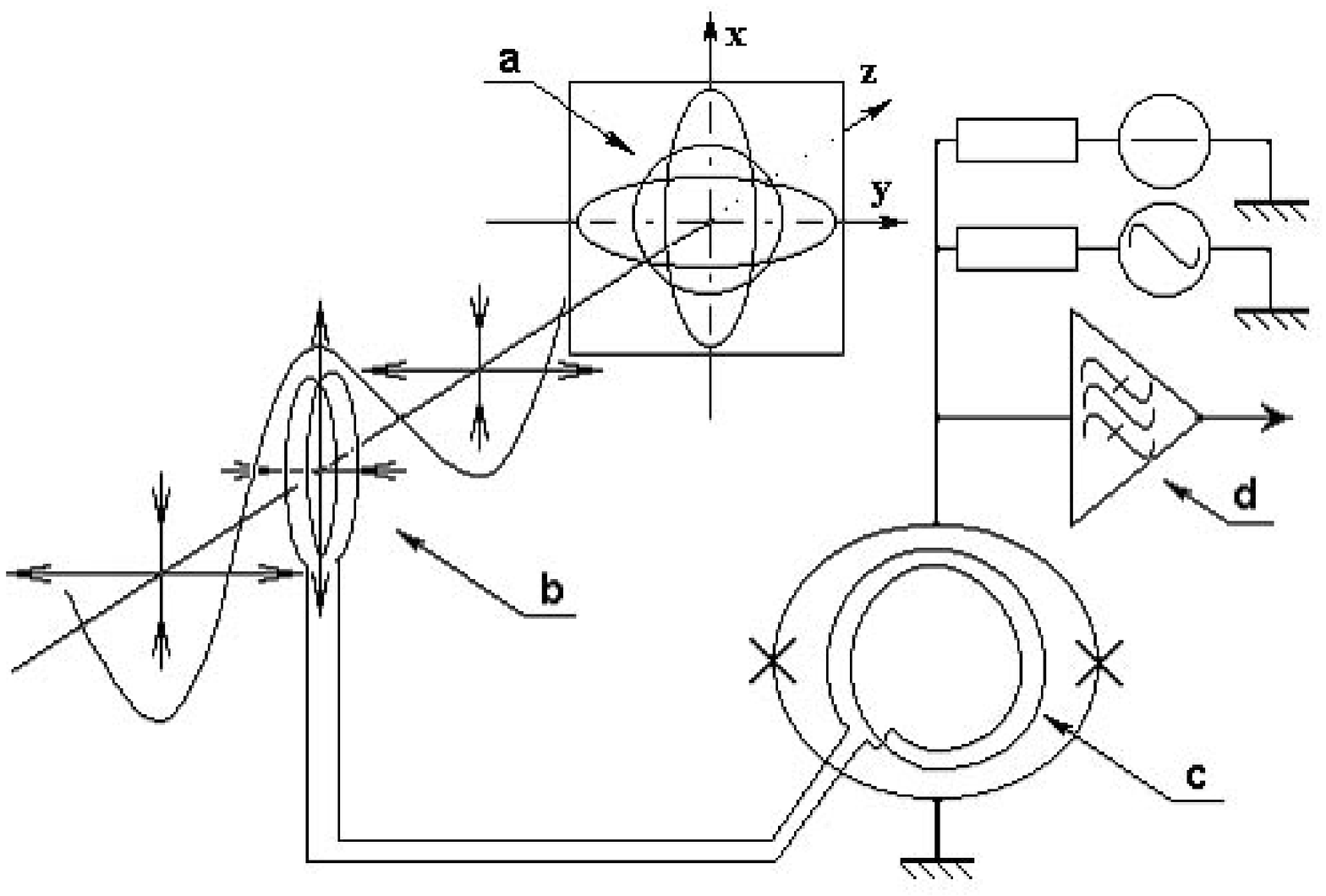}\par
  \vspace{6pt}
  \parbox{117mm}{\small {\itshape Figure 4.} The schematic diagram of the conversion
  of gravitational perturbations into the signals that are detectable by SQUID:
  \ $\mathbf{^a}$\,the distortion (``breathing'') of a circle, depicted inside a plane that is
  perpendicular to the wave vector; \ $\mathbf{^b}$\,the stretching of the effective area of the
  conversion coil in flux superconductive transformers; \ $\mathbf{^c}$\,the coupling coil that
  leads magnetic signals in DC-SQUID; \ $\mathbf{^d}$\,the input device from the electronic
  composition of DC-SQUID.}\par
\end{center}
\vspace{36pt}

Such jump of the elasticity must generate the gigantic wave
reflection on the boundary of matter-vacuum. By the same reason,
the process of the energy transfer from GW into mechanical
oscillations is done much better than the inverse process. This
very inconvertibility at the similar transformation (when others
the less fundamental noise sources will be obviated) will not
eventually allow us to bring nearer the noise temperature of the
GW detector to absolute zero and makes meaningless the next step –
the application of the quantum squeezing in the recording system
evidently. The way out from such fundamental deadlock is the
application of described above recording systems without direct
energy conversion. In these ones the conversion irreversibility is
lacking. The direct conversion of a gravitational perturbation
into a phase response that SQUID is able to trap may be carried
out by a superconductive transformer of the magnetic flux
(fig.~4).  The transformer is a closed superconductive circuit,
which is composed from a couple of coils, each of them has an
inductance value which is very close to the other's one: the
conversion coil (fig.~4, b) and the coupling coil (fig.~4, c). The
{\it coupling} (or {\it loop}) coil leads changes of the magnetic
flux, generated by GW, to the ``sensitive'' element of SQUID. The
{\it conversion} coil is put in such a way as to arrange the plane
of its own rings to be parallel to the wave vector of perturbed
GW. In such a configuration (fig.~5), according to Stockes
theorem, any variation of the metric tensor $g_{ij}$ may cause
some change of the trapped magnetic flux that flows through the
conversion coil circuit (i.\,e. through the plane XOZ,
fig.~5):
$$\delta\Phi = \delta\oint g_{ij}\,A^i\,dr^j \approx \oint\delta g_{ij}\,A^i\,dr^j,$$
where $A^i$ are the components of vector potential $\mathbf A$,
$dr^j$ are the components of radius-vector $\mathbf r$; the
integral is taken along the whole closed contour of
superconductive conversion coil, which lies in the plane XOZ.
Therefore, any perturbation of the metric causes some small
increment of the magnetic flux, whose (approximately) half will
fall into the transmission coil. It is well known that when
passing through the ``ordinary'' three-dimensional space, a plane
gravitational wave incidents so that the circle, represented in
the plane XOZ (fig.~5, a), which is perpendicular to the wave
vector, is periodically stretched in one direction and squeezed in
another direction (as it were ``breathing'') while it leaves the
area of the visible ellipse constant all the time. The
stretching/shrinkage of the ellipse's axes, as well as of all
linear dimensions inside the plane XOZ which is perpendicular to
the wave propagation direction, leads to the change of the
conversion coil's working area (while the wave vector is parallel
to the planes of its rings), and this fact allows us to estimate
the flux increment without need of calculating any circulation
integral. The relative stretching/shrinkage of the ellipse's sizes
is estimated approximately by the variation of the metric tensor.
Consequently, the relative increments of the conversion coil's
effective area and the magnetic flux are estimated by that one.
Therefore, if in the transformer one freezes a magnetic flux \
$\Phi_C \simeq 20\,mWb \approx 10^{13}\Phi_0$, \ where \ $\Phi_0 =
2.07\!\times\!10^{-15}\,Wb$ \ is the flux quantum, then a
gravitational wave whose amplitude of the metric tensor's
oscillation is in the order of \ $|\,\delta g_{ij}\,| \simeq
10^{-20}$ \ will give at the input of SQUID a flux increment of
order \ $\Delta\Phi \approx \Phi_C|\,\delta g_{ij}\,| \simeq
10^{-7}\Phi_0$, \ that complies with the limit resolution of a
modern two-stage DC-SQUID, which is in the order of \
$0.1\,Hz$~\cite{Cla}.

\vspace{18pt}
\begin{center}
  \includegraphics[width=117mm,keepaspectratio]{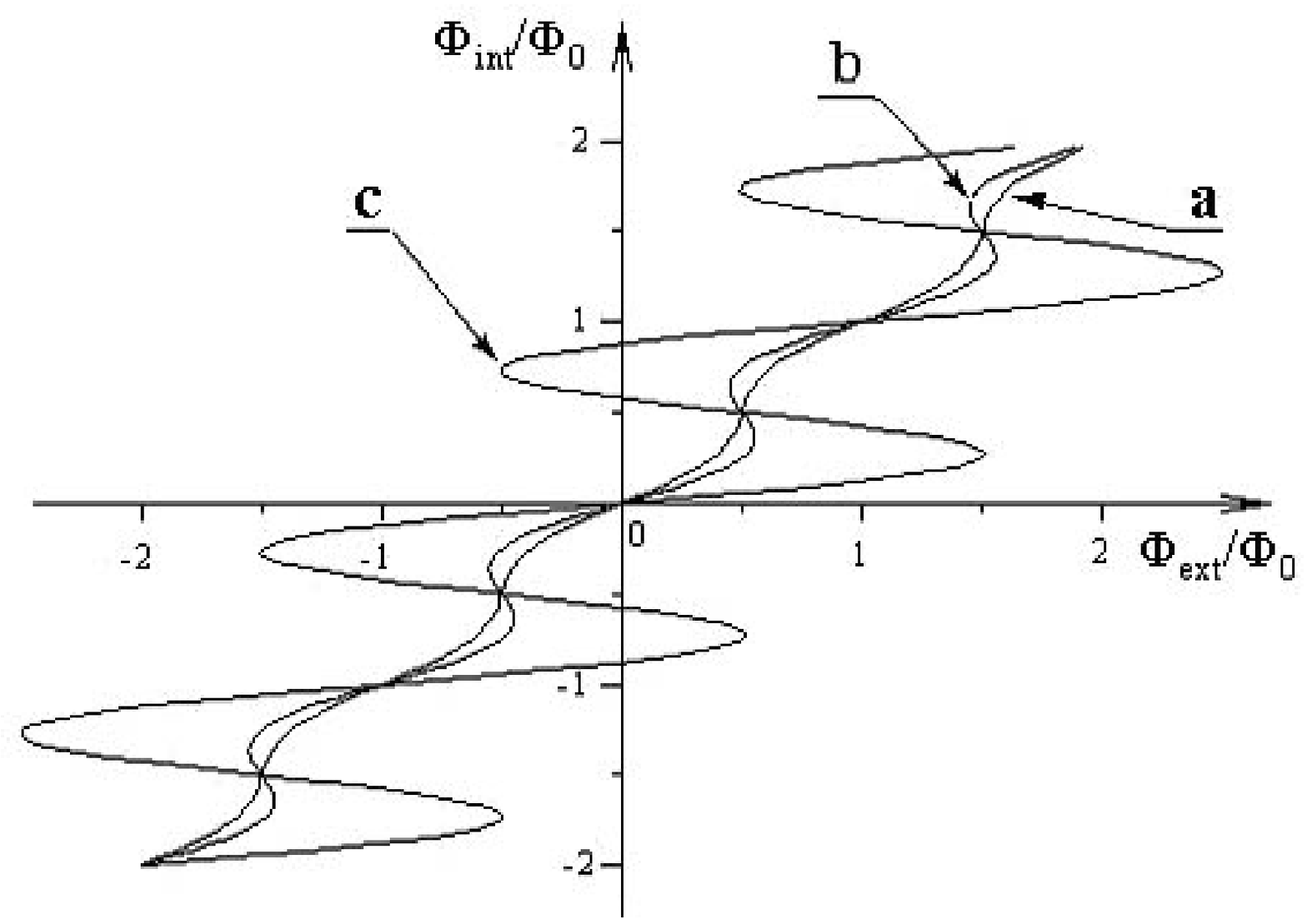}\par
  \vspace{6pt}
  \parbox{117mm}{\small {\itshape Figure 5.} The plots that represent the dependence of the
  internal magnetic flux ($\Phi_{int}$, as the ordinate axis) inside a superconductive loop,
  switching into the circuit the Josephson tunnel transmission and functioning as an active
  transformer: \ $\mathbf{^a}$\,singlevalued anhysteresis loop; \ $\mathbf{^b}$\,multiplevalued
  hysteresis loop; \ $\mathbf{^c}$\,``utrahysteresis'' loop.}\par
\end{center}
\vspace{36pt}

According to our estimations, the design of terra-hertzian
anhysteresis UHF-SQUID's using the quantum squeezing (fitting with
Josephson plasmic oscillations) of coherent states is promising to
increase the sensitivity by more than three orders (i.\,e. with \
$\Delta\Phi \approx 10^{-10}\,\Phi_0$). At the same time it
becomes available the detection of GW with the amplitude of order
\ $|\,\delta g_{ij}\,| \simeq 10^{-23}$ \ in the band of $1\,Hz$.
It is necessary to note that in the quoted estimations we have not
yet taken in to account the signal losses, unavoidable when
adjusting the SQUID's input circuit with the transmission coil
(these losses are estimated to be of order 1–2). Based on the
principle of the reversibility of linear electrodynamic systems,
it is possible to offer a circuit for directly converting the
variations of a gravitational field into phase responses, by means
of utilizing an active superconductive transformer for the fluxes
which includes the Josephson tunnel transition. In that case flux
and phase are dependent on each other, but the linearity is still
ensured by the smallness of the variations. The functions of the
conversion and connection coils in an active transformer are
integrated on the principle ``two in one'', and the tunnel
transition is put into the gap of the superconductive loop. The
described structure is essentially a ``superhysteresis'' HF-SQUID,
i.\,e.~~$LI_C \gg \Phi_0$, \ where $L$ is the coil inductance,
$I_C$ is the critical current intensity of Josephson tunnel
transmission. The wide range of the multivalued hysteresis branch
($\pm LI_C$) is in compliance with the small value of the
derivative \ $d\Phi_{int}/d\Phi_{ext} \approx \Phi_0/(LI_C)$ \
outside the vicinities of the points with \ $\Phi_{int} =
(n+\frac{1}{2})\,\Phi_0/2$ \,(in the linear domain); here
$\Phi_{int}$ is the internal magnetic flux through a loop,
$\Phi_{ext}$ is the external flux, and $n$ is an integer. The
internal flux $\Phi_{int}$ (represented by ordinate axis in
fig.~5) is determined by the phase difference in Josephson
transmission, \ $\Phi_{int} = \Phi_0$. \ In response to the
distortion of the metric in the closed circuit of an active
transformer, it is brought about an increment of just the phase
diference, that leads to the change of the internal
flux~\cite{Bry}. The variation of the last, according to the
principle of reversibility, has to result in an increment of the
external flux which is proportional to \
$(d\Phi_{int}/d\Phi_{ext})^{-1}$. \ Consequently,
$$\partial\Phi_{ext} \approx (LI_C/\Phi_0)\,\Phi_{int}\,|\,\delta g_{ij}\,| \approx (LI_C/\Phi_0)^2\,\Phi_0\,|\,\delta g_{ij}\,|.$$
Otherwise, in terms of the flux quantum, the response of the
external magnetic field on the dynamical variation of the metric
is given by the expression \ $\Delta\Phi_{ext}/\Phi_0 \approx
(LI_C/\Phi_0)^2\,|\,\delta g_{ij}\,|$.

In order to estimate the limit resolution of the set-up in terms
of metric instead of \ $\Delta\Phi_{ext}/\Phi_0$, \ it is
necessary to set the resolution by the flux through SQUID, which
is registering the external field, generated by an active
transformer in response to the space-time distortion \ $|\,\delta
g_{ij}\,| \approx (\Delta\Phi/\Phi_0)/(LI_C/\Phi_0)^2$. \ If by
way of $LI_C$ to take \ $20\,\mu Wb = 10^9\,\Phi_0$, \ then even
at the sensitivity of SQUID \ $\Delta\Phi/\Phi_0 \simeq
10^{-6}\,Hz^{-1/2}$ \ the set-up will be able to detect GW with
the record-breaking small amplitude \ $|\,\delta g_{ij}\,| \simeq
10^{-24}\,Hz^{-1/2}$.

\end{document}